\documentclass[aps,prl,reprint,superscriptaddress]{revtex4-1}
\usepackage{graphicx}
\usepackage{dcolumn}
\usepackage{bm} 
\usepackage{braket} 
\usepackage[usenames, dvipsnames]{color}
\usepackage{color}

\begin{document}


\title{Disordered Route to the Coulomb Quantum Spin Liquid: Random Transverse Fields on Spin Ice in Pr$_2$Zr$_2$O$_7$}


\author{J.-J. Wen}
\affiliation{Institute for Quantum Matter and Department of Physics and Astronomy, The Johns Hopkins University, Baltimore, MD 21218, USA}
\affiliation{Department of Applied Physics, Stanford University, Stanford, CA 94305, USA}
\affiliation{Stanford Institute for Materials and Energy Sciences, SLAC National Accelerator Laboratory, 2575 Sand Hill Road, Menlo Park, CA 94025, USA}

\author{S. M. Koohpayeh}
\affiliation{Institute for Quantum Matter and Department of Physics and Astronomy, The Johns Hopkins University, Baltimore, MD 21218, USA}

\author{K. A. Ross}
\affiliation{Institute for Quantum Matter and Department of Physics and Astronomy, The Johns Hopkins University, Baltimore, MD 21218, USA}
\affiliation{NIST Center for Neutron Research, National Institute of Standards and Technology, Gaithersburg, MD 20899, USA}

\author{B. A. Trump}
\affiliation{Department of Chemistry, The Johns Hopkins University, Baltimore, MD 21218, USA}

\author{T. M. McQueen}
\affiliation{Institute for Quantum Matter and Department of Physics and Astronomy, The Johns Hopkins University, Baltimore, MD 21218, USA}
\affiliation{Department of Chemistry, The Johns Hopkins University, Baltimore, MD 21218, USA}
\affiliation{Department of Materials Science and Engineering, The Johns Hopkins University, Baltimore, MD 21218, USA}

\author{K. Kimura}
\affiliation{Institute for Solid State Physics (ISSP), University of Tokyo, Kashiwa, Chiba 277-8581, Japan}
\affiliation{Division of Materials Physics, Graduate School of Engineering Science, Osaka University, Toyonaka, Osaka 560-8531, Japan}

\author{S. Nakatsuji}
\affiliation{Institute for Solid State Physics (ISSP), University of Tokyo, Kashiwa, Chiba 277-8581, Japan}
\affiliation{CREST, Japan Science and Technology Agency, Kawaguchi, Saitama 332-0012, Japan}

\author{Y. Qiu}
\affiliation{NIST Center for Neutron Research, National Institute of Standards and Technology, Gaithersburg, MD 20899, USA}

\author{D. M. Pajerowski}
\affiliation{NIST Center for Neutron Research, National Institute of Standards and Technology, Gaithersburg, MD 20899, USA}

\author{J. R. D. Copley}
\affiliation{NIST Center for Neutron Research, National Institute of Standards and Technology, Gaithersburg, MD 20899, USA}

\author{C. L. Broholm}
\affiliation{Institute for Quantum Matter and Department of Physics and Astronomy, The Johns Hopkins University, Baltimore, MD 21218, USA}
\affiliation{NIST Center for Neutron Research, National Institute of Standards and Technology, Gaithersburg, MD 20899, USA}
\affiliation{Department of Materials Science and Engineering, The Johns Hopkins University, Baltimore, MD 21218, USA}



\date{\today}

\begin{abstract}
Inelastic neutron scattering reveals a broad continuum of excitations in Pr$_2$Zr$_2$O$_7$, the temperature and magnetic field dependence of which indicate a continuous distribution of quenched transverse fields ($\Delta$) acting on the non-Kramers Pr$^{3+}$ crystal field ground state doublets. Spin-ice correlations are apparent within 0.2 meV of the Zeeman energy. A random phase approximation provides an excellent account of the data with a transverse field distribution $\rho(\Delta)\propto (\Delta^2+\Gamma^2)^{-1}$ where $ \Gamma=0.27(1)$~meV. Established during high temperature synthesis due to an underlying structural instability, it appears disorder in Pr$_2$Zr$_2$O$_7$ actually induces a quantum spin liquid.  
\end{abstract}

\pacs{}

\maketitle


Instead of a discrete set of states that satisfy all interactions, frustrated spin systems support a high density of low energy states from which novel collective phenomena may emerge at temperatures ($T$) well below the bare interaction strengths.\cite{lacroix2011introduction} A prominent example is quantum spin ice (QSI).\cite{[][{, and references therein.}]Gingras2014} By introducing quantum spin fluctuations to classical spin ice through transverse inter-spin interactions, it has been proposed that a quantum spin liquid (QSL) phase with gapless photon-like excitations can be realized.\cite{Gingras2014}

Several materials have been examined in the search for QSI including $\rm Yb_2Ti_2O_7$ and $\rm Tb_2Ti_2O_7$, but so far there is no experimental evidence for salient features such as low energy electrodynamics.\cite{Gingras2014} Instead unanticipated features have been discovered including a very strong dependence of physical properties on sample quality. In a recent study of $\rm Tb_{2+x}Ti_{2-x}O_{7+y}$ it was found that a change in the Tb/Ti molar ratio as small as 0.005 can tune the samples between an ordered and a disordered phase.\cite{PhysRevB.87.060408} While such sensitivity is a distinguishing feature of systems with a high density of low energy states, there is so far no clear understanding of the microscopic mechanisms involved. Can this be explained in terms of small changes of the exchange interactions in the pseudo-spin-1/2 Hamiltonian\cite{Gingras2014} for materials near phase boundaries or are there new pieces of the puzzle yet to be discovered? Further insight into these questions will not only help clarify the complicated magneto-structural properties of specific materials, but may guide the broader search for QSL materials.
 
In this paper we show quenched structural disorder acts as a transverse field on the non-Kramers Pr$^{3+}$ ion in Pr$_2$Zr$_2$O$_7$ (PZO) and in competition with exchange interactions induces a spatially correlated and disordered singlet ground state. A previous neutron study of PZO revealed weak diffuse elastic scattering with pinch points indicative of spin-ice correlations.\cite{Kimura2013} Here we show magnetic excitations in PZO are composed of two parts: a lower energy regime that is driven by inter-spin correlations, and a momentum transfer ($\bm{q}$) independent higher energy part driven by quenched transverse fields. A nearest neighbor spin ice model augmented by random transverse fields provides an excellent account of these observations and points to the realization of a newly proposed QSL\cite{savary2016} in PZO.

 Single crystalline PZO samples were prepared by the floating zone method. Growth conditions were optimized to produce homogeneous stoichiometric crystals.\cite{Koohpayeh2014291} Five (9.5 g) and three (8.8 g) crystals were co-aligned for measurements in $(HHL)$ and $(HK0)$ planes respectively. Neutron measurements were carried out on the Disk Chopper Spectrometer\cite{Copley2003477} at the NIST Center for Neutron Research. Unless otherwise noted, data were collected with 3.3 meV incident neutron energy ($E_i$). Absolute units for the scattering cross section were obtained by normalization to the (220) nuclear Bragg peak. 

 \begin{figure}[t]
 \includegraphics[scale=0.38]{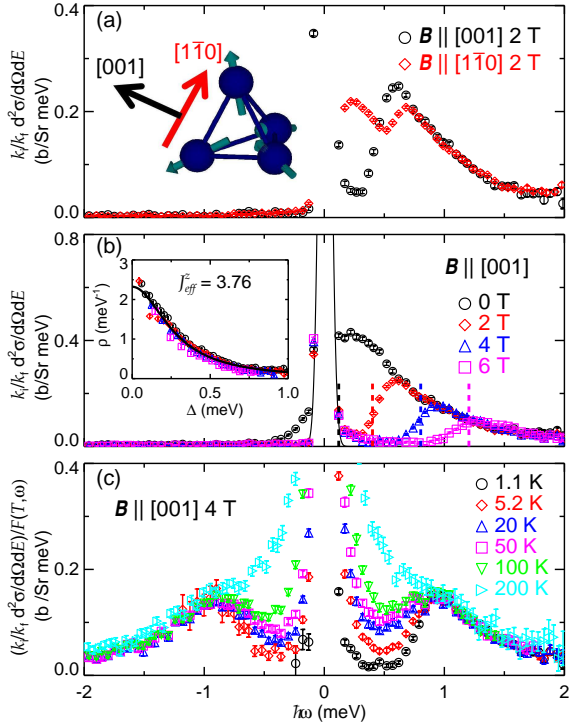}%
 \caption{\label{fig:1}$\bm{q}$-averaged spectra for $0.5$ \AA $^{-1} \leq  |\bm{q}| \leq 2.2$ \AA $^{-1}$ in the corresponding scattering planes. (a) Spectra for $\bm{B} \parallel [001], T = 1.4$ K and $\bm{B} \parallel [1\bar10], T = 0.2$ K. $|\bm{B}|= 2$ T. Inset shows the relative orientations of the applied fields and the spins. (b) Field dependent spectra for $\bm{B} \parallel [001]$ at $T = 1.4$ K. Inset shows the transverse field distribution $\rho(\Delta)$, and the solid line shows the half Lorentzian fit. Dashed vertical lines show the corresponding energies above which data were used to extract $\rho(\Delta)$, and the solid line shows the elastic scattering contribution at 0~T. (c) $T$-dependent spectra for $\bm{B} \parallel [001],$ 4 T. The data was divided by $F(T,\omega)$ of Eq.~(1). Error bars represent one standard deviation. The energy resolution at the elastic line is 0.11~meV.}
 \end{figure}

$\bm{q}$-averaged inelastic neutron scattering (INS) data provide an overview of the excitation spectrum (Fig.~\ref{fig:1}). At $T=1.4$~K and magnetic field $\bm{B}=0$~T, the spectrum takes the form of a broad peak that extends to at least 2 meV, consistent with previous measurements.\cite{Kimura2013} This spectrum is an order of magnitude broader than estimates for the magnetic exchange constant extracted from bulk thermodynamic data.\cite{Kimura2013} 
 
 Fig. ~\ref{fig:1}(a) shows INS measured under [$1\bar{1}0$] and [001] field directions  which have 2 and 1 components respectively. The $\braket{111}$ Ising anisotropy in PZO\cite{Kimura2013} implies that for $\bm{B} \parallel [1\bar10]$, spins are partitioned into two sets of chains with different field projections along the easy axes, while for $\bm{B} \parallel [001]$, the angles from easy $\braket{111}$ type directions to the field axis are identical [Fig.~\ref{fig:1}(a) inset]. The INS in Fig. ~\ref{fig:1}(a) thus reveal that these excitations are influenced by the field component along the easy axis and are of magnetic origin.  

The INS measured for different $\bm{B} \parallel [001]$ [Fig. ~\ref{fig:1}(b)] show an unusual field dependence. For para- and ferro- magnetic materials the application of a magnetic field shifts the INS intensities to higher energies. In PZO there is only a loss of spectral weight at lower energies while the spectral weight at higher energies is unaffected. As implied by the total moment sum-rule\cite{lovesey1986theory}, we find the missing spectral weight shifts to the elastic line. 

 The thermal evolution of the spectrum is essential for understanding the underlying physics. Fig.~\ref{fig:1}(c) shows the spectra measured with $\bm{B}$ = 4 T $\parallel$ [001] at temperatures from 1.1 K to 200 K. The $\bm{q}$-averaged INS were divided by a thermal factor $F(T,\omega)$
 \begin{eqnarray}
 F(T,\omega)=[1+e^{-\beta\hbar\omega}+e^{-\beta\hbar\omega/2}Z'(T)]^{-1}.
 \end{eqnarray}
Here $\beta=1/k_\mathrm{B}T$ and $Z'=\sum_{E'}e^{-\beta E'}$ where $E'$ runs through all CEF levels above the nominal ground state doublet (GSD).\cite{Kimura2013} If the excitations correspond to transitions between the CEF GSD that are split by an energy gap of $\hbar\omega$, the normalized spectra at different temperatures should fall on a single curve.\cite{PhysRevLett.94.177201} This pertains for $|\hbar\omega| \geq 0.9$ meV and $T \leq 200$~K, which indicates a continuous quenched distribution of doublet splittings is responsible for the higher energy part of the spectrum.

 \begin{figure}[t]
 \includegraphics[scale=0.38]{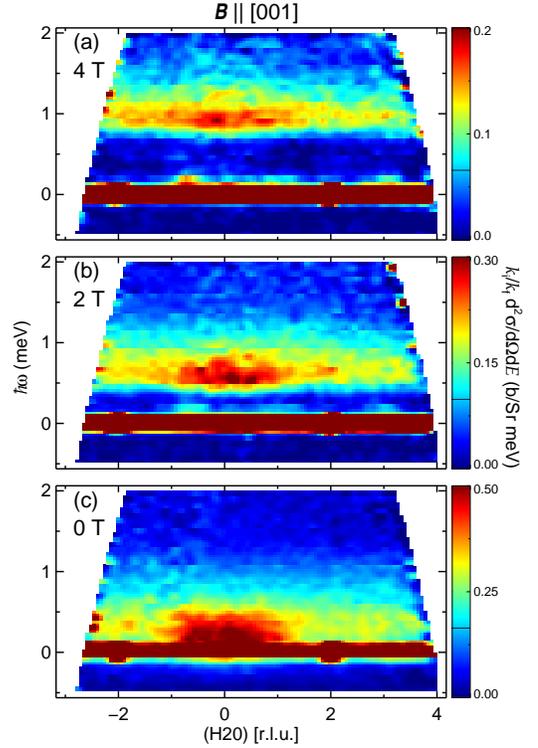}%
 \caption{\label{fig:2}Field dependent $\bm{q}$-$\omega$ slices for $\bm{B} \parallel [001], T = 1.4$ K. Data were integrated along $(0K0)$ for 1 (r.l.u.) $\leq K \leq$ 3 (r.l.u.). }
 \end{figure}

A closer look at the spectrum with $\bm{q}$-resolution provides crucial information regarding spin-spin correlations. Fig.~\ref{fig:2} shows $\bm{q}-\omega$ cuts along the $(H20)$ direction for various $\bm{B} \parallel [001]$. A broad spectrum is observed throughout $\bm{q}$ space that moves to higher energies with increasing field with no discernible dispersion. This contrasts with $\rm Tb_2Ti_2O_7$\cite{PhysRevLett.96.177201} and $\rm Yb_2Ti_2O_7$\cite{Ross2011} where resonant spin wave modes develop at high fields. The main $\bm{q}$-dependent aspect of the PZO data is intensity modulation at the lower energy edge with a maximum near $(020)$. This modulation exceeds that associated with the magnetic form factor in the $\bm{q}$ range probed, so it reflects inter-site correlations. The modulation diminishes at higher energies, consistent with Fig.~\ref{fig:1}(c) where the high energy part of the spectrum is a local CEF excitation.

\begin{figure}[t]
 \includegraphics[scale=0.38]{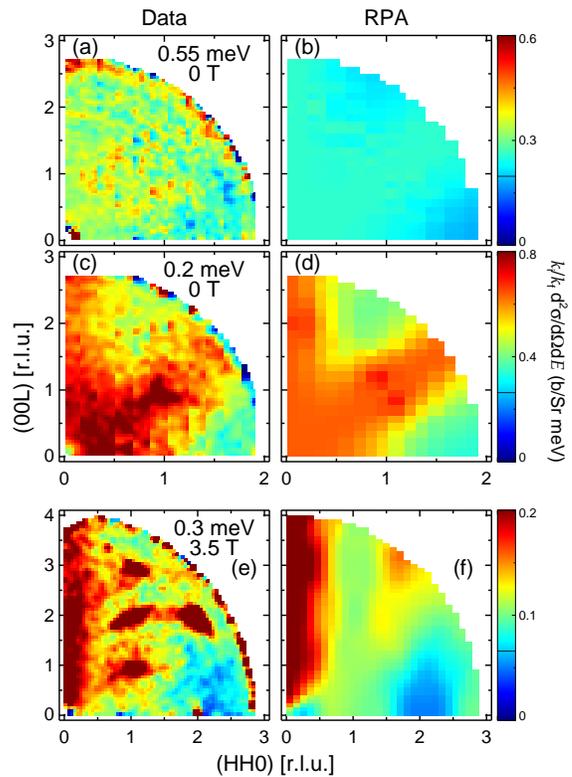}%
 \caption{\label{fig:3}$\bm{q}$-maps in $(HHL)$ plane. Data were folded into the first quadrant for optimal statistics. (a) and (c) show measurements with $E_i=1.7$~meV at 0.05 K, 0 T with energy integration of [0.3, 0.8] meV and [0.1, 0.3] meV respectively. (e) shows data with $E_i=3.3$~meV at 0.2 K, $\bm{B}=3.5$~T $\parallel [1\bar10]$ with energy integration of [0.1, 0.5] meV. The energy resolutions at the elastic line are 0.04 meV ($E_i=1.7$~meV) and 0.11 meV ($E_i=3.3$~meV) respectively. (b), (d), and (f) show corresponding RPA calculations.}
 \end{figure}

More information regarding spatial spin correlations is extracted by constant energy slices through the data. Fig.~\ref{fig:3} shows such slices covering the $(HHL)$ plane. Figs.~\ref{fig:3}(a)(c) show measurements at $\bm{B}=0$ T. While no modulation is observed at high energies [Fig.~\ref{fig:3}(a)], a star-fish-like intensity modulation [Fig.~\ref{fig:3}(c)] is observed at low energies that is consistent with previous measurements.\cite{Kimura2013} By applying $\bm{B} \parallel [1\bar10]$, the low energy modulation evolves into a single rod along $(00L)$ that is characteristic of low dimensional correlations [Fig.~\ref{fig:3}(e)]. As previously mentioned, a magnetic field along $[1\bar10]$ separates the pyrochlore lattice into two distinct sets of chains where the external field is perfectly transverse to - or has a component along - the local easy axis. Correspondingly this low energy scattering can be associated with Pr chains perpendicular to both the [001] and magnetic field directions. For those spins the magnetic field is perpendicular to the easy axis. In Fig.~\ref{fig:3}(e), the intensity maxima at (111), (112), (113), and (222) are instrumental spurions that result from leakage beyond the elastic channel of strong nuclear and field induced magnetic elastic scattering.

To sum up the experimental observations presented so far: (1) The higher energy part of the INS is consistent with scattering from a continuum of heterogeneously split doublets, the distribution of which is continuous and $T$-independent up to 200 K. (2) Inter-site correlations are apparent for the lower energy part of the INS which is modulated in $\bm{q}$ space.

To understand the data we start with a description of the single ion part of the problem. Since the relevant energy range is well below the first excited CEF level ($\sim 9.5$~meV\cite{Kimura2013}), it is a good approximation to work in the reduced Hilbert space of the GSD. There the total spin operator $\bm{J}$ in the local frame takes the form of $J^x = 0, J^y=0, J^z = J^z_{eff}\sigma^z$, with $z$ the local $\braket{111}$ directions. Here $\sigma$ are the Pauli matrices and $J^z_{eff}$ is the effective spin size of the GSD. The zero field splitting of the doublet can be described by a transverse field, $\Delta (\geq 0)$.\cite{Baker156} Thus without loss of generality the single ion part of the spin Hamiltonian can be written as

\begin{eqnarray}
\mathcal{H}_0 = \sum_i (-g\mu_\mathrm{B}\bm{B}\cdot\bm{J}_i +\Delta_i \sigma_i^x),
\label{eq:1}
\end{eqnarray}
here $g$ is the Land\'e $g$-factor. The first term describes the Zeeman coupling and the second term accounts for the fixed splitting of each doublet. We fix $J^z_{eff}=3.76$ to the value determined from the GSD wavefunction\cite{Kimura2013}, which should be considered as a sample averaged value. The distribution from site to site of energy gaps for the split doublets is described in terms of a distribution function $\rho(\Delta)$. For a given $\rho(\Delta)$ the corresponding INS can be calculated from $\mathcal{H}_0$ using Fermi's golden rule\cite{squires2012}[see the supplementary information (SI)]. Thus from the experimental spectra we can extract $\rho(\Delta)$ and if $\mathcal{H}_0$ is a good approximation for PZO, $\rho(\Delta)$ extracted from data acquired at different fields should be consistent. This analysis was carried out for the field dependent spectra in Fig.~\ref{fig:1}(b). For zero field, data points with $\hbar\omega \ge 0.1$~meV were used to avoid influences from the elastic line. For finite fields, data points with $\hbar\omega$ larger than the Zeeman gap ($\frac{2}{\sqrt{3}}g\mu_\mathrm{B}|\bm{B}|J^z_{eff}$) were used. A common normalization factor was applied to ensure unity normalization: $\int_0^\infty \!\rho(\Delta)\, \mathrm{d}\Delta = 1$. As shown in the inset to Fig.~\ref{fig:1}(b), $\rho(\Delta)$ extracted at different fields fall onto a single curve, which can be described by a ``half'' Lorentzian: $\rho(\Delta)=2\Gamma[\pi(\Delta^2+\Gamma^2)]^{-1}\Theta(\Delta)$ where $ \Gamma=0.27(1)$~meV. We expect $\rho(\Delta)$ to be accurate for $\Delta$ larger than the exchange energy scale ($\sim0.1$~meV). The success of a single ion description and the associated continuous distribution function $\rho(\Delta)$ indicates that Pr in PZO is subject to a broad range of static random transverse fields extending at least up to $1$~meV.

 \begin{figure}[t]
 \includegraphics[scale=0.4]{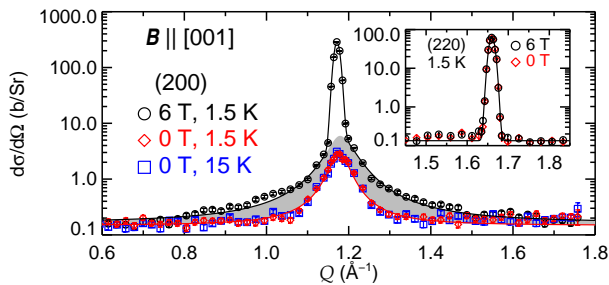}%
 \caption{\label{fig:4}Elastic longitudinal cuts for the (200) forbidden nuclear peak. Solid lines are fits to the data. Shaded area shows the broad field induced elastic magnetic scattering. Inset shows corresponding cuts through the (220) nuclear peak. }
 \end{figure}

Elastic scattering measurements provide further information regarding the spatial inhomogeneity of $\Delta_i$ in PZO. For $\bm{B} \parallel [001]$ all spins are subject to the same local field, so that a homogeneous sample should be uniformly magnetized and support sharp magnetic Bragg diffraction. Fig.~\ref{fig:4} shows that in addition to a sharp field induced magnetic Bragg peak at (200), there is an underlying broad component (shaded). This indicates inhomogeneous magnetization that is correlated over $\sim22.8$~\AA (2.1 cubic lattice spacings)\cite{[][{Correlation length is estimated by 2/FWHM, where FWHM is full width at half maximum determined from fitting.}]corrlen}, which can be considered a measure of the spatial profile of the transverse field distribution. In comparison, the (220) nuclear Bragg peak remains sharp at all fields. There is a broad elastic peak at (200) in zero field (Fig.~\ref{fig:4}) that persists to temperatures well beyond the inter-site interaction scale and  may reflect the underlying structural disorder that is associated with the transverse field distribution.    

While $\mathcal{H}_0$ provides a good description of the higher energy part of the spectrum, inter-site interactions are important at lower energies. We consider the minimal interacting Hamiltonian, which adds ferromagnetic near neighbor interactions to the random transverse field single ion terms:
\begin{eqnarray}
\mathcal{H}=\mathcal{J} \sum_{\braket{i,j}}\bm{J}_i\cdot\bm{J}_j+ \mathcal{H}_0,
\label{eq:three}
\end{eqnarray}
here $\braket{i,j}$ run over distinct nearest neighbors. If the activation energy inferred from AC susceptibility is the energy of a single ice-rule-violating tetrahedron\cite{Kimura2013} then $\mathcal{J}$ for PZO can be estimated to be $\mathcal{J}=0.015$ meV. We analyze $\mathcal H$ through the random phase approximation (RPA).\cite{jensen1991} To account for the random spatial distribution of $\Delta_i$ in $\mathcal{H}_0$ we first calculate the RPA susceptibility $\chi_{\rm{RPA}}(\bm{q},\omega,\Delta)$ assuming a constant $\Delta$, and then approximate the sample averaged susceptibility as a weighted average: $\bar{\chi}_{\rm{RPA}}(\bm{q},\omega)=\int_{0}^\infty \!\rho(\Delta)\chi_{\rm{RPA}}(\bm{q},\omega,\Delta)\,\mathrm{d}\Delta$. This approach is expected to work well for small $\Delta$ where $\rho(\Delta)$ is large, while for large and rare values of $\Delta$ it is expected to overestimate correlations since at least in a simplified uncorrelated model of disorder a spin with large $\Delta$ is less likely to have neighbors with similar $\Delta$. The INS is then calculated from $\bar{\chi}_{\rm{RPA}}$ through the fluctuation-dissipation theorem.\cite{lovesey1986theory} All the RPA calculations were scaled by a common factor to compare to the experimental data. Figs.~\ref{fig:3} (b), (d), and (f) show this simple RPA model can describe the experimental results.

Having established $\mathcal{H}$ as consistent with INS from PZO, we consider the physical origin of $\rho(\Delta)$. Pr$^{3+}$ being a non-Kramers ion, $\Delta_i$ can arise from any deviation in the coordinating environment that breaks the $D_{3d}$ point group symmetry. X-ray diffraction from single crystalline PZO down to $110$~K shows no sign of a structural distortion from the ideal pyrochlore structure, which places strict limits on any cooperative Jahn-Teller distortion (see SI). The deviation from pyrochlore symmetry is thus expected to be local and break translational symmetry. A weak, broad peak was indeed observed at the forbidden (200) position (Fig.~\ref{fig:4}). The temperature independence of this peak (probed for temperatures between 1.5 K and 15 K) indicates quenched structural disorder, which could be related to the Pr off-centering observed in polycrystalline $\rm Pr_{2+x}Zr_{2-x}O_{7-x/2}$.\cite{Koohpayeh2014291}  A Lorentzian fit to the peak yields a correlation length of 39.9 \AA, which is comparable to the 22.8 \AA~correlation length extracted for the field induced magnetization. Note that multiple scattering is not relevant for the broad peak observed here.\cite{PhysRevB.92.024110}

Might a disorder induced singlet ground state be relevant to other non-Kramers pyrochlore materials? Structural disorder can act as a random transverse field on any non-Kramers ions ($\rm Pr^{3+}$, $\rm Pm^{3+}$, $\rm Sm^{2+}$, $\rm Eu^{3+}$, $\rm Tb^{3+}$, $\rm Ho^{3+}$, $\rm Tm^{3+}$). When exchange interactions are sufficiently strong and non-frustrated, magnetic order might still be possible. However in weakly interacting and frustrated materials it seems reasonable to anticipate a disordered singlet ground state. Inhomogeneous doublet splitting was previously reported in $\rm Pr_2Ru_2O_7$\cite{PhysRevLett.94.177201}, so PZO is not a singular example. The phenomenon may also be relevant to understanding $\rm Tb_2Ti_2O_7$. There the first exited CEF level lies at 1.5~meV\cite{PhysRevB.62.6496} (compared to 9.5~meV for PZO), so it partakes in the low-$T$ physics. Judging from INS in the literature\cite{PhysRevB.86.174403,Gaulin2011}, any transverse field distribution in $\rm Tb_2Ti_2O_7$ is significantly narrower than for PZO (perhaps 0.2~meV wide compared to $1$~meV for PZO). Both compounds display an extreme sensitivity to metal stoichiometry, which in $\rm Pr_{2+x}Zr_{2-x}O_{7-x/2}$ has been linked to Pr off-centering.\cite{Koohpayeh2014291} This suggests that in $\rm Tb_{2+x}Ti_{2-x}O_{7+y}$, x may also control the strength of a random transverse field that quenches its magnetism.\cite{PhysRevB.87.060408,Bonville2011} If this is the case then pressure induced magnetic order in $\rm Tb_2Ti_2O_7$ could be associated with varying the transverse field distribution.\cite{PhysRevLett.93.187204}  

Clearly, quenched transverse fields strongly influence low-$T$ magnetism for PZO. Sites with $\Delta\gg \mathcal{J}(J^z_{eff})^2\sim 2$~K will be essentially non-magnetic and act as dilution to the magnetic pyrochlore lattice. This can be expected to suppress the nuclear Schottky anomaly in specific heat and increase specific heat in a temperature range set by $\Gamma$ as previously observed.\cite{Kimura2013,Koohpayeh2014291}  Also, the spatial landscape of $\Delta_i$ is expected to constrain the development of spin-ice correlations, leading to a reduced correlation length evident in broadened pinch points.\cite{Kimura2013} In the presence of strong static transverse fields a reliable estimate of the transverse exchange interactions is difficult to obtain. Suffice it to say that  $\mathcal{H}$ which contains only the longitudinal Ising interactions, provides an acceptable account of the data. 

QSI is expected to be stable to all local perturbations despite being gapless.\cite{Gingras2014} A recent theoretical study of $\mathcal{H}$ shows random transverse fields induce quantum entanglement and two distinct QSLs.\cite{savary2016} The gapless nature of the spectrum [Fig.~\ref{fig:1}(b)] and the spin-ice-like correlations reported here (Fig.~\ref{fig:3}) and in previous studies\cite{Kimura2013} preclude a trivial paramagnet and point to a disorder induced QSL in PZO.  

Given the importance of random transverse fields that we have demonstrated in $\rm Pr_2Zr_2O_7$ and the potential to access distinct QSLs\cite{savary2016} by controlling disorder through a combination of chemical doping\cite{Koohpayeh2014291} and hydrostatic pressure, disorder in non-Kramers frustrated magnets now appears to present a rich area for the experimental realization and exploration of solid state quantum entanglement. 

\begin{acknowledgments}
The work at IQM was supported by the US Department of Energy, office of Basic Energy Sciences, Division of Materials Sciences and Engineering under grant DE-FG02-08ER46544. This work is partially supported by CREST, Japan Science and Technology Agency, by Grants-in-Aid for Scientific Research (16H02209), by Grants-in-Aids for Scientific Research on Innovative Areas (15H05882, 15H05883) and Program for Advancing Strategic International Networks to Accelerate the Circulation of Talented Researchers (No. R2604) from the Japanese Society for the Promotion of Science. This work utilized facilities supported in part by the National Science Foundation under Agreement No. DMR-1508249. J. W. acknowledges the support at Stanford and SLAC by the U.S. Department of Energy (DOE), Office of Science, Basic Energy Sciences, Materials Sciences and Engineering Division, under Contract No. DE-AC02-76SF00515.
\end{acknowledgments}

\textit{Note added.} After completion of our work, we became aware of an independent work by Petit \textit{et al.} on the same compound, which suggests transverse interactions could be important in PZO.\cite{PhysRevB.94.165153}


\begin{thebibliography}{22}%
\makeatletter
\providecommand \@ifxundefined [1]{%
 \@ifx{#1\undefined}
}%
\providecommand \@ifnum [1]{%
 \ifnum #1\expandafter \@firstoftwo
 \else \expandafter \@secondoftwo
 \fi
}%
\providecommand \@ifx [1]{%
 \ifx #1\expandafter \@firstoftwo
 \else \expandafter \@secondoftwo
 \fi
}%
\providecommand \natexlab [1]{#1}%
\providecommand \enquote  [1]{``#1''}%
\providecommand \bibnamefont  [1]{#1}%
\providecommand \bibfnamefont [1]{#1}%
\providecommand \citenamefont [1]{#1}%
\providecommand \href@noop [0]{\@secondoftwo}%
\providecommand \href [0]{\begingroup \@sanitize@url \@href}%
\providecommand \@href[1]{\@@startlink{#1}\@@href}%
\providecommand \@@href[1]{\endgroup#1\@@endlink}%
\providecommand \@sanitize@url [0]{\catcode `\\12\catcode `\$12\catcode
  `\&12\catcode `\#12\catcode `\^12\catcode `\_12\catcode `\%12\relax}%
\providecommand \@@startlink[1]{}%
\providecommand \@@endlink[0]{}%
\providecommand \url  [0]{\begingroup\@sanitize@url \@url }%
\providecommand \@url [1]{\endgroup\@href {#1}{\urlprefix }}%
\providecommand \urlprefix  [0]{URL }%
\providecommand \Eprint [0]{\href }%
\providecommand \doibase [0]{http://dx.doi.org/}%
\providecommand \selectlanguage [0]{\@gobble}%
\providecommand \bibinfo  [0]{\@secondoftwo}%
\providecommand \bibfield  [0]{\@secondoftwo}%
\providecommand \translation [1]{[#1]}%
\providecommand \BibitemOpen [0]{}%
\providecommand \bibitemStop [0]{}%
\providecommand \bibitemNoStop [0]{.\EOS\space}%
\providecommand \EOS [0]{\spacefactor3000\relax}%
\providecommand \BibitemShut  [1]{\csname bibitem#1\endcsname}%
\let\auto@bib@innerbib\@empty
\bibitem [{\citenamefont {Lacroix}\ \emph {et~al.}(2011)\citenamefont
  {Lacroix}, \citenamefont {Mendels},\ and\ \citenamefont
  {Mila}}]{lacroix2011introduction}%
  \BibitemOpen
  \bibfield  {author} {\bibinfo {author} {\bibfnamefont {C.}~\bibnamefont
  {Lacroix}}, \bibinfo {author} {\bibfnamefont {P.}~\bibnamefont {Mendels}}, \
  and\ \bibinfo {author} {\bibfnamefont {F.}~\bibnamefont {Mila}},\ }\href@noop
  {} {\emph {\bibinfo {title} {Introduction to Frustrated Magnetism: Materials,
  Experiments, Theory}}},\ Springer Series in Solid-State Sciences\ (\bibinfo
  {publisher} {Springer Berlin Heidelberg},\ \bibinfo {year}
  {2011})\BibitemShut {NoStop}%
\bibitem [{\citenamefont {Gingras}\ and\ \citenamefont
  {McClarty}(2014)}]{Gingras2014}%
  \BibitemOpen
  \bibfield  {author} {\bibinfo {author} {\bibfnamefont {M.~J.~P.}\
  \bibnamefont {Gingras}}\ and\ \bibinfo {author} {\bibfnamefont {P.~A.}\
  \bibnamefont {McClarty}},\ }\href
  {http://stacks.iop.org/0034-4885/77/i=5/a=056501} {\bibfield  {journal}
  {\bibinfo  {journal} {Reports on Progress in Physics}\ }\textbf {\bibinfo
  {volume} {77}},\ \bibinfo {pages} {056501} (\bibinfo {year}
  {2014})}\BibitemShut {NoStop}%
\bibitem [{\citenamefont {Taniguchi}\ \emph {et~al.}(2013)\citenamefont
  {Taniguchi}, \citenamefont {Kadowaki}, \citenamefont {Takatsu}, \citenamefont
  {F\aa{}k}, \citenamefont {Ollivier}, \citenamefont {Yamazaki}, \citenamefont
  {Sato}, \citenamefont {Yoshizawa}, \citenamefont {Shimura}, \citenamefont
  {Sakakibara}, \citenamefont {Hong}, \citenamefont {Goto}, \citenamefont
  {Yaraskavitch},\ and\ \citenamefont {Kycia}}]{PhysRevB.87.060408}%
  \BibitemOpen
  \bibfield  {author} {\bibinfo {author} {\bibfnamefont {T.}~\bibnamefont
  {Taniguchi}}, \bibinfo {author} {\bibfnamefont {H.}~\bibnamefont {Kadowaki}},
  \bibinfo {author} {\bibfnamefont {H.}~\bibnamefont {Takatsu}}, \bibinfo
  {author} {\bibfnamefont {B.}~\bibnamefont {F\aa{}k}}, \bibinfo {author}
  {\bibfnamefont {J.}~\bibnamefont {Ollivier}}, \bibinfo {author}
  {\bibfnamefont {T.}~\bibnamefont {Yamazaki}}, \bibinfo {author}
  {\bibfnamefont {T.~J.}\ \bibnamefont {Sato}}, \bibinfo {author}
  {\bibfnamefont {H.}~\bibnamefont {Yoshizawa}}, \bibinfo {author}
  {\bibfnamefont {Y.}~\bibnamefont {Shimura}}, \bibinfo {author} {\bibfnamefont
  {T.}~\bibnamefont {Sakakibara}}, \bibinfo {author} {\bibfnamefont
  {T.}~\bibnamefont {Hong}}, \bibinfo {author} {\bibfnamefont {K.}~\bibnamefont
  {Goto}}, \bibinfo {author} {\bibfnamefont {L.~R.}\ \bibnamefont
  {Yaraskavitch}}, \ and\ \bibinfo {author} {\bibfnamefont {J.~B.}\
  \bibnamefont {Kycia}},\ }\href {\doibase 10.1103/PhysRevB.87.060408}
  {\bibfield  {journal} {\bibinfo  {journal} {Phys. Rev. B}\ }\textbf {\bibinfo
  {volume} {87}},\ \bibinfo {pages} {060408} (\bibinfo {year}
  {2013})}\BibitemShut {NoStop}%
\bibitem [{\citenamefont {Kimura}\ \emph {et~al.}(2013)\citenamefont {Kimura},
  \citenamefont {Nakatsuji}, \citenamefont {Wen}, \citenamefont {Broholm},
  \citenamefont {Stone}, \citenamefont {Nishibori},\ and\ \citenamefont
  {Sawa}}]{Kimura2013}%
  \BibitemOpen
  \bibfield  {author} {\bibinfo {author} {\bibfnamefont {K.}~\bibnamefont
  {Kimura}}, \bibinfo {author} {\bibfnamefont {S.}~\bibnamefont {Nakatsuji}},
  \bibinfo {author} {\bibfnamefont {J.-J.}\ \bibnamefont {Wen}}, \bibinfo
  {author} {\bibfnamefont {C.}~\bibnamefont {Broholm}}, \bibinfo {author}
  {\bibfnamefont {M.~B.}\ \bibnamefont {Stone}}, \bibinfo {author}
  {\bibfnamefont {E.}~\bibnamefont {Nishibori}}, \ and\ \bibinfo {author}
  {\bibfnamefont {H.}~\bibnamefont {Sawa}},\ }\href {\doibase
  10.1038/ncomms2914} {\bibfield  {journal} {\bibinfo  {journal} {Nature
  communications}\ }\textbf {\bibinfo {volume} {4}},\ \bibinfo {pages} {1934}
  (\bibinfo {year} {2013})}\BibitemShut {NoStop}%
\bibitem [{\citenamefont {Savary}\ and\ \citenamefont
  {Balents}(2016)}]{savary2016}%
  \BibitemOpen
  \bibfield  {author} {\bibinfo {author} {\bibfnamefont {L.}~\bibnamefont
  {Savary}}\ and\ \bibinfo {author} {\bibfnamefont {L.}~\bibnamefont
  {Balents}},\ }\href@noop {} {\bibfield  {journal} {\bibinfo  {journal} {ArXiv
  e-prints}\ } (\bibinfo {year} {2016})},\ \Eprint
  {http://arxiv.org/abs/1604.04630} {arXiv:1604.04630 [cond-mat.str-el]}
  \BibitemShut {NoStop}%
\bibitem [{\citenamefont {Koohpayeh}\ \emph {et~al.}(2014)\citenamefont
  {Koohpayeh}, \citenamefont {Wen}, \citenamefont {Trump}, \citenamefont
  {Broholm},\ and\ \citenamefont {McQueen}}]{Koohpayeh2014291}%
  \BibitemOpen
  \bibfield  {author} {\bibinfo {author} {\bibfnamefont {S.}~\bibnamefont
  {Koohpayeh}}, \bibinfo {author} {\bibfnamefont {J.-J.}\ \bibnamefont {Wen}},
  \bibinfo {author} {\bibfnamefont {B.}~\bibnamefont {Trump}}, \bibinfo
  {author} {\bibfnamefont {C.}~\bibnamefont {Broholm}}, \ and\ \bibinfo
  {author} {\bibfnamefont {T.}~\bibnamefont {McQueen}},\ }\href {\doibase
  http://dx.doi.org/10.1016/j.jcrysgro.2014.06.037} {\bibfield  {journal}
  {\bibinfo  {journal} {Journal of Crystal Growth}\ }\textbf {\bibinfo {volume}
  {402}},\ \bibinfo {pages} {291 } (\bibinfo {year} {2014})}\BibitemShut
  {NoStop}%
\bibitem [{\citenamefont {Copley}\ and\ \citenamefont
  {Cook}(2003)}]{Copley2003477}%
  \BibitemOpen
  \bibfield  {author} {\bibinfo {author} {\bibfnamefont {J.}~\bibnamefont
  {Copley}}\ and\ \bibinfo {author} {\bibfnamefont {J.}~\bibnamefont {Cook}},\
  }\href {\doibase http://dx.doi.org/10.1016/S0301-0104(03)00124-1} {\bibfield
  {journal} {\bibinfo  {journal} {Chem. Phys.}\ }\textbf {\bibinfo {volume}
  {292}},\ \bibinfo {pages} {477 } (\bibinfo {year} {2003})}\BibitemShut
  {NoStop}%
\bibitem [{\citenamefont {Lovesey}(1986)}]{lovesey1986theory}%
  \BibitemOpen
  \bibfield  {author} {\bibinfo {author} {\bibfnamefont {S.}~\bibnamefont
  {Lovesey}},\ }\href@noop {} {\emph {\bibinfo {title} {Theory of Neutron
  Scattering from Condensed Matter}}},\ International series of monographs on
  physics\ (\bibinfo  {publisher} {Clarendon Press},\ \bibinfo {year}
  {1986})\BibitemShut {NoStop}%
\bibitem [{\citenamefont {van Duijn}\ \emph {et~al.}(2005)\citenamefont {van
  Duijn}, \citenamefont {Kim}, \citenamefont {Hur}, \citenamefont {Adroja},
  \citenamefont {Adams}, \citenamefont {Huang}, \citenamefont {Jaime},
  \citenamefont {Cheong}, \citenamefont {Broholm},\ and\ \citenamefont
  {Perring}}]{PhysRevLett.94.177201}%
  \BibitemOpen
  \bibfield  {author} {\bibinfo {author} {\bibfnamefont {J.}~\bibnamefont {van
  Duijn}}, \bibinfo {author} {\bibfnamefont {K.~H.}\ \bibnamefont {Kim}},
  \bibinfo {author} {\bibfnamefont {N.}~\bibnamefont {Hur}}, \bibinfo {author}
  {\bibfnamefont {D.}~\bibnamefont {Adroja}}, \bibinfo {author} {\bibfnamefont
  {M.~A.}\ \bibnamefont {Adams}}, \bibinfo {author} {\bibfnamefont {Q.~Z.}\
  \bibnamefont {Huang}}, \bibinfo {author} {\bibfnamefont {M.}~\bibnamefont
  {Jaime}}, \bibinfo {author} {\bibfnamefont {S.-W.}\ \bibnamefont {Cheong}},
  \bibinfo {author} {\bibfnamefont {C.}~\bibnamefont {Broholm}}, \ and\
  \bibinfo {author} {\bibfnamefont {T.~G.}\ \bibnamefont {Perring}},\ }\href
  {\doibase 10.1103/PhysRevLett.94.177201} {\bibfield  {journal} {\bibinfo
  {journal} {Phys. Rev. Lett.}\ }\textbf {\bibinfo {volume} {94}},\ \bibinfo
  {pages} {177201} (\bibinfo {year} {2005})}\BibitemShut {NoStop}%
\bibitem [{\citenamefont {Rule}\ \emph {et~al.}(2006)\citenamefont {Rule},
  \citenamefont {Ruff}, \citenamefont {Gaulin}, \citenamefont {Dunsiger},
  \citenamefont {Gardner}, \citenamefont {Clancy}, \citenamefont {Lewis},
  \citenamefont {Dabkowska}, \citenamefont {Mirebeau}, \citenamefont {Manuel},
  \citenamefont {Qiu},\ and\ \citenamefont {Copley}}]{PhysRevLett.96.177201}%
  \BibitemOpen
  \bibfield  {author} {\bibinfo {author} {\bibfnamefont {K.~C.}\ \bibnamefont
  {Rule}}, \bibinfo {author} {\bibfnamefont {J.~P.~C.}\ \bibnamefont {Ruff}},
  \bibinfo {author} {\bibfnamefont {B.~D.}\ \bibnamefont {Gaulin}}, \bibinfo
  {author} {\bibfnamefont {S.~R.}\ \bibnamefont {Dunsiger}}, \bibinfo {author}
  {\bibfnamefont {J.~S.}\ \bibnamefont {Gardner}}, \bibinfo {author}
  {\bibfnamefont {J.~P.}\ \bibnamefont {Clancy}}, \bibinfo {author}
  {\bibfnamefont {M.~J.}\ \bibnamefont {Lewis}}, \bibinfo {author}
  {\bibfnamefont {H.~A.}\ \bibnamefont {Dabkowska}}, \bibinfo {author}
  {\bibfnamefont {I.}~\bibnamefont {Mirebeau}}, \bibinfo {author}
  {\bibfnamefont {P.}~\bibnamefont {Manuel}}, \bibinfo {author} {\bibfnamefont
  {Y.}~\bibnamefont {Qiu}}, \ and\ \bibinfo {author} {\bibfnamefont {J.~R.~D.}\
  \bibnamefont {Copley}},\ }\href {\doibase 10.1103/PhysRevLett.96.177201}
  {\bibfield  {journal} {\bibinfo  {journal} {Phys. Rev. Lett.}\ }\textbf
  {\bibinfo {volume} {96}},\ \bibinfo {pages} {177201} (\bibinfo {year}
  {2006})}\BibitemShut {NoStop}%
\bibitem [{\citenamefont {Ross}\ \emph {et~al.}(2011)\citenamefont {Ross},
  \citenamefont {Savary}, \citenamefont {Gaulin},\ and\ \citenamefont
  {Balents}}]{Ross2011}%
  \BibitemOpen
  \bibfield  {author} {\bibinfo {author} {\bibfnamefont {K.~A.}\ \bibnamefont
  {Ross}}, \bibinfo {author} {\bibfnamefont {L.}~\bibnamefont {Savary}},
  \bibinfo {author} {\bibfnamefont {B.~D.}\ \bibnamefont {Gaulin}}, \ and\
  \bibinfo {author} {\bibfnamefont {L.}~\bibnamefont {Balents}},\ }\href
  {\doibase 10.1103/PhysRevX.1.021002} {\bibfield  {journal} {\bibinfo
  {journal} {Phys. Rev. X}\ }\textbf {\bibinfo {volume} {1}},\ \bibinfo {pages}
  {021002} (\bibinfo {year} {2011})}\BibitemShut {NoStop}%
\bibitem [{\citenamefont {Baker}\ and\ \citenamefont
  {Bleaney}(1958)}]{Baker156}%
  \BibitemOpen
  \bibfield  {author} {\bibinfo {author} {\bibfnamefont {J.~M.}\ \bibnamefont
  {Baker}}\ and\ \bibinfo {author} {\bibfnamefont {B.}~\bibnamefont
  {Bleaney}},\ }\href {\doibase 10.1098/rspa.1958.0074} {\bibfield  {journal}
  {\bibinfo  {journal} {Proceedings of the Royal Society of London A:
  Mathematical, Physical and Engineering Sciences}\ }\textbf {\bibinfo {volume}
  {245}},\ \bibinfo {pages} {156} (\bibinfo {year} {1958})}\BibitemShut
  {NoStop}%
\bibitem [{\citenamefont {Squires}(2012)}]{squires2012}%
  \BibitemOpen
  \bibfield  {author} {\bibinfo {author} {\bibfnamefont {G.}~\bibnamefont
  {Squires}},\ }\href@noop {} {\emph {\bibinfo {title} {Introduction to the
  Theory of Thermal Neutron Scattering}}}\ (\bibinfo  {publisher} {Cambridge
  University Press},\ \bibinfo {year} {2012})\BibitemShut {NoStop}%
\bibitem [{cor()}]{corrlen}%
  \BibitemOpen
  \href@noop {} {\ }\BibitemShut {NoStop}%
\bibitem [{\citenamefont {Jensen}\ and\ \citenamefont
  {Mackintosh}(1991)}]{jensen1991}%
  \BibitemOpen
  \bibfield  {author} {\bibinfo {author} {\bibfnamefont {J.}~\bibnamefont
  {Jensen}}\ and\ \bibinfo {author} {\bibfnamefont {A.~R.}\ \bibnamefont
  {Mackintosh}},\ }\href@noop {} {\emph {\bibinfo {title} {Rare Earth
  Magnetism: Structures and Excitations}}}\ (\bibinfo  {publisher} {Clarendon
  Press},\ \bibinfo {year} {1991})\BibitemShut {NoStop}%
\bibitem [{\citenamefont {Baroudi}\ \emph {et~al.}(2015)\citenamefont
  {Baroudi}, \citenamefont {Gaulin}, \citenamefont {Lapidus}, \citenamefont
  {Gaudet},\ and\ \citenamefont {Cava}}]{PhysRevB.92.024110}%
  \BibitemOpen
  \bibfield  {author} {\bibinfo {author} {\bibfnamefont {K.}~\bibnamefont
  {Baroudi}}, \bibinfo {author} {\bibfnamefont {B.~D.}\ \bibnamefont {Gaulin}},
  \bibinfo {author} {\bibfnamefont {S.~H.}\ \bibnamefont {Lapidus}}, \bibinfo
  {author} {\bibfnamefont {J.}~\bibnamefont {Gaudet}}, \ and\ \bibinfo {author}
  {\bibfnamefont {R.~J.}\ \bibnamefont {Cava}},\ }\href {\doibase
  10.1103/PhysRevB.92.024110} {\bibfield  {journal} {\bibinfo  {journal} {Phys.
  Rev. B}\ }\textbf {\bibinfo {volume} {92}},\ \bibinfo {pages} {024110}
  (\bibinfo {year} {2015})}\BibitemShut {NoStop}%
\bibitem [{\citenamefont {Gingras}\ \emph {et~al.}(2000)\citenamefont
  {Gingras}, \citenamefont {den Hertog}, \citenamefont {Faucher}, \citenamefont
  {Gardner}, \citenamefont {Dunsiger}, \citenamefont {Chang}, \citenamefont
  {Gaulin}, \citenamefont {Raju},\ and\ \citenamefont
  {Greedan}}]{PhysRevB.62.6496}%
  \BibitemOpen
  \bibfield  {author} {\bibinfo {author} {\bibfnamefont {M.~J.~P.}\
  \bibnamefont {Gingras}}, \bibinfo {author} {\bibfnamefont {B.~C.}\
  \bibnamefont {den Hertog}}, \bibinfo {author} {\bibfnamefont
  {M.}~\bibnamefont {Faucher}}, \bibinfo {author} {\bibfnamefont {J.~S.}\
  \bibnamefont {Gardner}}, \bibinfo {author} {\bibfnamefont {S.~R.}\
  \bibnamefont {Dunsiger}}, \bibinfo {author} {\bibfnamefont {L.~J.}\
  \bibnamefont {Chang}}, \bibinfo {author} {\bibfnamefont {B.~D.}\ \bibnamefont
  {Gaulin}}, \bibinfo {author} {\bibfnamefont {N.~P.}\ \bibnamefont {Raju}}, \
  and\ \bibinfo {author} {\bibfnamefont {J.~E.}\ \bibnamefont {Greedan}},\
  }\href {\doibase 10.1103/PhysRevB.62.6496} {\bibfield  {journal} {\bibinfo
  {journal} {Phys. Rev. B}\ }\textbf {\bibinfo {volume} {62}},\ \bibinfo
  {pages} {6496} (\bibinfo {year} {2000})}\BibitemShut {NoStop}%
\bibitem [{\citenamefont {Petit}\ \emph {et~al.}(2012)\citenamefont {Petit},
  \citenamefont {Bonville}, \citenamefont {Robert}, \citenamefont {Decorse},\
  and\ \citenamefont {Mirebeau}}]{PhysRevB.86.174403}%
  \BibitemOpen
  \bibfield  {author} {\bibinfo {author} {\bibfnamefont {S.}~\bibnamefont
  {Petit}}, \bibinfo {author} {\bibfnamefont {P.}~\bibnamefont {Bonville}},
  \bibinfo {author} {\bibfnamefont {J.}~\bibnamefont {Robert}}, \bibinfo
  {author} {\bibfnamefont {C.}~\bibnamefont {Decorse}}, \ and\ \bibinfo
  {author} {\bibfnamefont {I.}~\bibnamefont {Mirebeau}},\ }\href {\doibase
  10.1103/PhysRevB.86.174403} {\bibfield  {journal} {\bibinfo  {journal} {Phys.
  Rev. B}\ }\textbf {\bibinfo {volume} {86}},\ \bibinfo {pages} {174403}
  (\bibinfo {year} {2012})}\BibitemShut {NoStop}%
\bibitem [{\citenamefont {Gaulin}\ \emph {et~al.}(2011)\citenamefont {Gaulin},
  \citenamefont {Gardner}, \citenamefont {McClarty},\ and\ \citenamefont
  {Gingras}}]{Gaulin2011}%
  \BibitemOpen
  \bibfield  {author} {\bibinfo {author} {\bibfnamefont {B.~D.}\ \bibnamefont
  {Gaulin}}, \bibinfo {author} {\bibfnamefont {J.~S.}\ \bibnamefont {Gardner}},
  \bibinfo {author} {\bibfnamefont {P.~A.}\ \bibnamefont {McClarty}}, \ and\
  \bibinfo {author} {\bibfnamefont {M.~J.~P.}\ \bibnamefont {Gingras}},\ }\href
  {\doibase 10.1103/PhysRevB.84.140402} {\bibfield  {journal} {\bibinfo
  {journal} {Phys. Rev. B}\ }\textbf {\bibinfo {volume} {84}},\ \bibinfo
  {pages} {140402} (\bibinfo {year} {2011})}\BibitemShut {NoStop}%
\bibitem [{\citenamefont {Bonville}\ \emph {et~al.}(2011)\citenamefont
  {Bonville}, \citenamefont {Mirebeau}, \citenamefont {Gukasov}, \citenamefont
  {Petit},\ and\ \citenamefont {Robert}}]{Bonville2011}%
  \BibitemOpen
  \bibfield  {author} {\bibinfo {author} {\bibfnamefont {P.}~\bibnamefont
  {Bonville}}, \bibinfo {author} {\bibfnamefont {I.}~\bibnamefont {Mirebeau}},
  \bibinfo {author} {\bibfnamefont {A.}~\bibnamefont {Gukasov}}, \bibinfo
  {author} {\bibfnamefont {S.}~\bibnamefont {Petit}}, \ and\ \bibinfo {author}
  {\bibfnamefont {J.}~\bibnamefont {Robert}},\ }\href {\doibase
  10.1103/PhysRevB.84.184409} {\bibfield  {journal} {\bibinfo  {journal} {Phys.
  Rev. B}\ }\textbf {\bibinfo {volume} {84}},\ \bibinfo {pages} {184409}
  (\bibinfo {year} {2011})}\BibitemShut {NoStop}%
\bibitem [{\citenamefont {Mirebeau}\ \emph {et~al.}(2004)\citenamefont
  {Mirebeau}, \citenamefont {Goncharenko}, \citenamefont {Dhalenne},\ and\
  \citenamefont {Revcolevschi}}]{PhysRevLett.93.187204}%
  \BibitemOpen
  \bibfield  {author} {\bibinfo {author} {\bibfnamefont {I.}~\bibnamefont
  {Mirebeau}}, \bibinfo {author} {\bibfnamefont {I.~N.}\ \bibnamefont
  {Goncharenko}}, \bibinfo {author} {\bibfnamefont {G.}~\bibnamefont
  {Dhalenne}}, \ and\ \bibinfo {author} {\bibfnamefont {A.}~\bibnamefont
  {Revcolevschi}},\ }\href {\doibase 10.1103/PhysRevLett.93.187204} {\bibfield
  {journal} {\bibinfo  {journal} {Phys. Rev. Lett.}\ }\textbf {\bibinfo
  {volume} {93}},\ \bibinfo {pages} {187204} (\bibinfo {year}
  {2004})}\BibitemShut {NoStop}%
\bibitem [{\citenamefont {Petit}\ \emph {et~al.}(2016)\citenamefont {Petit},
  \citenamefont {Lhotel}, \citenamefont {Guitteny}, \citenamefont {Florea},
  \citenamefont {Robert}, \citenamefont {Bonville}, \citenamefont {Mirebeau},
  \citenamefont {Ollivier}, \citenamefont {Mutka}, \citenamefont {Ressouche},
  \citenamefont {Decorse}, \citenamefont {Ciomaga~Hatnean},\ and\ \citenamefont
  {Balakrishnan}}]{PhysRevB.94.165153}%
  \BibitemOpen
  \bibfield  {author} {\bibinfo {author} {\bibfnamefont {S.}~\bibnamefont
  {Petit}}, \bibinfo {author} {\bibfnamefont {E.}~\bibnamefont {Lhotel}},
  \bibinfo {author} {\bibfnamefont {S.}~\bibnamefont {Guitteny}}, \bibinfo
  {author} {\bibfnamefont {O.}~\bibnamefont {Florea}}, \bibinfo {author}
  {\bibfnamefont {J.}~\bibnamefont {Robert}}, \bibinfo {author} {\bibfnamefont
  {P.}~\bibnamefont {Bonville}}, \bibinfo {author} {\bibfnamefont
  {I.}~\bibnamefont {Mirebeau}}, \bibinfo {author} {\bibfnamefont
  {J.}~\bibnamefont {Ollivier}}, \bibinfo {author} {\bibfnamefont
  {H.}~\bibnamefont {Mutka}}, \bibinfo {author} {\bibfnamefont
  {E.}~\bibnamefont {Ressouche}}, \bibinfo {author} {\bibfnamefont
  {C.}~\bibnamefont {Decorse}}, \bibinfo {author} {\bibfnamefont
  {M.}~\bibnamefont {Ciomaga~Hatnean}}, \ and\ \bibinfo {author} {\bibfnamefont
  {G.}~\bibnamefont {Balakrishnan}},\ }\href {\doibase
  10.1103/PhysRevB.94.165153} {\bibfield  {journal} {\bibinfo  {journal} {Phys.
  Rev. B}\ }\textbf {\bibinfo {volume} {94}},\ \bibinfo {pages} {165153}
  (\bibinfo {year} {2016})}\BibitemShut {NoStop}%
\end{thebibliography}
%

\end{document}